\documentclass{jkas}

\def\year{2014} % publication year
\def\month{May} % publication month
\def\volume{00} % journal volume
\def\issue{0} % journal issue
\def\beginpage{1} % first page of article
\def\endpage{99} % last page of article
\def\received{February 30, 2014} % date paper was received by JKAS
\def\accepted{February 31, 2014} % date of acceptance
\date{Received \received; accepted \accepted}

\usepackage{placeins}

\title{
The automatic calibration of Korean VLBI Network data 
}

\author[1]{Jeffrey~A.~Hodgson}
\author[1,2]{Sang-Sung~Lee}
\author[1]{Guang-Yao~Zhao}
\author[1]{Juan-Carlos~Algaba}
\author[1]{Youngjoo~Yun}
\author[1]{Taehyun~Jung}
\author[1,2]{Do-Young~Byun}
\affil[1]{Korea Astronomy and Space Science Institute, 776 Daedeokdae-ro, Yuseong-gu, Daejeon 34055, Korea; \email{jhodgson@kasi.re.kr}}
\affil[2]{University of Science and Technology, 217 Gajeong-ro, Yuseong-gu, Daejeon, 34113, Korea}
\def\corrauthor{
J.A. Hodgson
}

\def\runningauthor{
J.A. Hodgson
}

\def\runningtitle{
Automatic KVN Calibration
}

\def\keywords{
KVN: pipeline: calibration
}

\def\abstracttext{
The calibration of Very Long Baseline Interferometry (VLBI) data has long been a time consuming process. The Korean VLBI Network (KVN) is a simple array consisting of three identical antennas.
Because four frequencies are observed simultaneously, phase solutions can be transferred from lower frequencies to higher frequencies in order to improve phase coherence and hence sensitivity at higher frequencies. Due to the homogeneous nature of the array, the KVN is also well suited for automatic calibration. In this paper we describe the automatic calibration of single-polarisation KVN data using the KVN Pipeline and comparing the results against VLBI data that has been manually reduced. We find that the pipelined data using phase transfer produces better results than a manually reduced dataset not using the phase transfer. Additionally we compared the pipeline results with a manually reduced phase-transferred dataset and found the results to be identical.
}

\begin{document}
\jkashead %% set title, authors, abstract, etc.

%%%%%%%%%%%%%%%%%%%%%%%%%%%%%%%%%%%%%%%%%%%%%%%%%%%%%%%%%%%%%%%%%%%%%
%%% BEGIN MAIN TEXT HERE %%%%%%%%%%%%%%%%%%%%%%%%%%%%%%%%%%%%%%%%%%%%
%%%%%%%%%%%%%%%%%%%%%%%%%%%%%%%%%%%%%%%%%%%%%%%%%%%%%%%%%%%%%%%%%%%%%

\section{Introduction\label{sec:intro}}

The calibration and analysis of Very Long Baseline Interferometry (VLBI) data has long been a time consuming process. However, with increases in computing power and hardware, it is becoming increasingly feasible to largely automate the process. The Korean VLBI Network (KVN) is an homogeneous radio array comprising of three antennas capable of simultaneously observing at 14\,mm, 7\,mm, 3.5\,mm and 2.3\,mm. The homogeneity of the array and the simultaneous multi-frequency capabilities allow for the transfer of phase solutions from lower frequencies to higher frequencies, make the KVN a suitable network to implement the automatic calibration of VLBI data.

% \begin{table*}[t]
% \caption{Overview of observations \label{tab:obs}}
% \centering
% \begin{tabular}{ccccccc}
% \toprule
% Code & Date & Length & Polarisation & Data rate & Wavelengths & Stations \\
%      &      &  [hours] &            &   [Gbps]        &   [mm]\\
% \midrule
% iMOGABA32 &  01-03-2016  &  24  & LCP & 1 & 15,7,3,2 & KY, KU, KT   \\
% \bottomrule
% \end{tabular}
% \tabnote{LCP: Left-hand circular polarisation.  Station codes: Yonsei (KY), Ulsan (KU), Tamna (KT).}
% \end{table*}

Arrays such as the KVN can produce as much as four datasets simultaneously, creating a large workload for those reducing the data. Due to this and other factors, there is a large backlog of non-reduced data, not just on the KVN but on most VLBI networks. For this reason, there is a large need for either fully automating or at least partially automatising the reduction process. The technical means for producing a fully automatic pipeline for single-polarisation continuum data are already developed with the largest constraints being typically logistical. A station should ideally provide calibration information (e.g. system temperature, antenna temperature and gain information) and also a monitoring log of the telescope, VLBI recorder, VLBI back-end and receiver. This information should allow for data to be automatically flagged. In many current VLBI experiments this information is unavailable or difficult to acquire. The KVN however largely provides this information. In this paper we describe a fully automatic pipeline for the KVN.

The European VLBI Network (EVN) is a large consortium of observatories that collaborate to perform regular VLBI experiments \citep{evn}. Two pipelines have been developed to aid the calibration of EVN data. Before 2006, a pipeline was written as a special procedure within the Astronomical Image Processing System (AIPS) \citep{aips}. The pipeline broadly followed the calibration procedure described in Section \ref{sec:calibration}, but was not completely automated - although it greatly speed up the calibration process \citep{evnpipe1}. A more recent pipeline also implemented in AIPS is the task \emph{VLBARUN}, which applied phase and amplitude calibration to Very Long Baseline Array (VLBA) data before preceding to produce a simple image. A newer pipeline was developed for the automated reduction of EVN data, written in the Python AIPS wrapper, ParselTongue and provides largely similar functionality to the earlier pipeline except while being more robust and easier to use \citep{parsel}. 

This paper describes the automatic reduction software (pipeline) of an example KVN dataset and compares the results of the KVN Pipeline with that of the same dataset reduced in the manual way. In Section \ref{sec:observations}, we describe the observations that were used to test the KVN Pipeline, in Section \ref{sec:calibration}, we describe the traditional approach to VLBI calibration, in Section \ref{sec:KVNpipe} we describe the operation of the KVN Pipeline, in Section \ref{sec:discussion} we discuss the results and compare them against the traditionally reduced VLBI data and in Section \ref{sec:con} we present our conclusions.

\section{Observations\label{sec:observations}}

\begin{figure*}[t]
\centering
\includegraphics[width=0.6\textwidth]{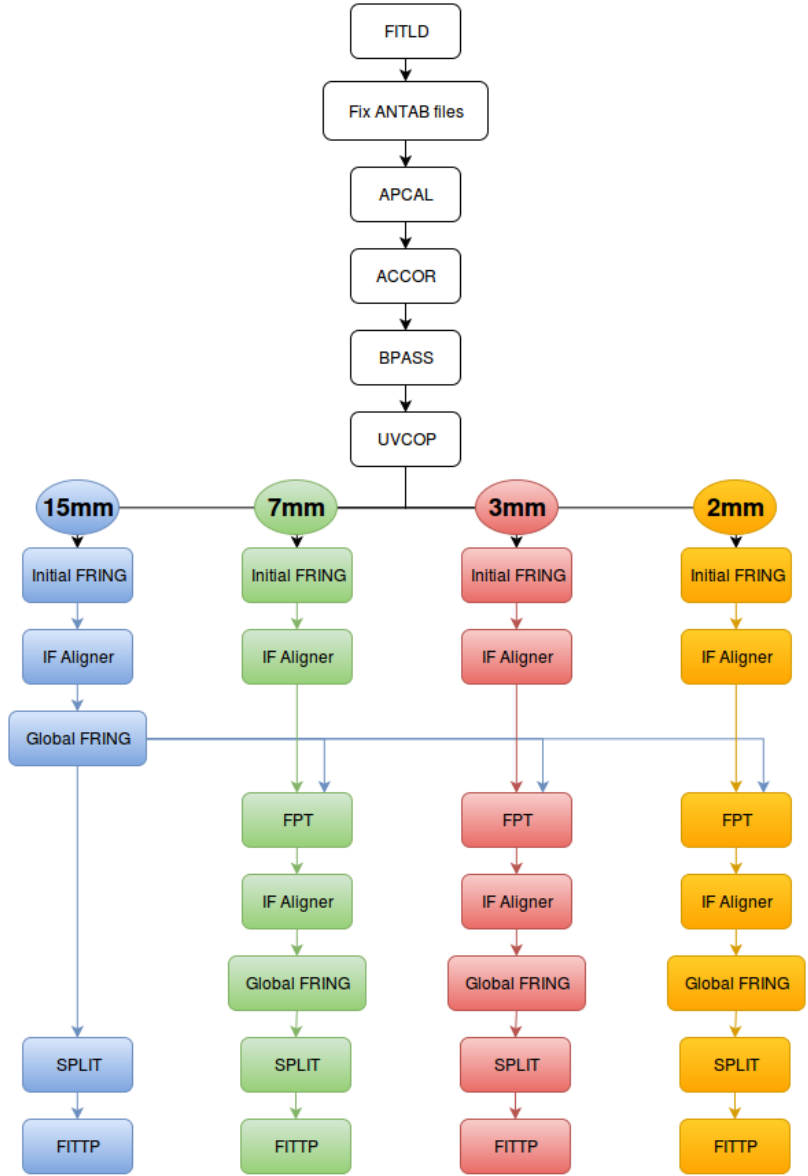}
\caption{A flow chart describing the calibration pipeline for the KVN.\label{fig:flowchart}}
\end{figure*}

The KVN is a telescope that consists of three antennas within South Korea in Seoul (Yonsei) (KY), Ulsan (KU) and Jeju Island (Tamna) (KT) \citep{kvn}. In order to describe the pipeline and analyse its performance, we selected a recent dataset observed as part of the interferometric MOnitoring of GAmma-ray Bright AGN (iMOGABA) program. This program has been monitoring $\gamma$-ray bright AGN using the KVN at 21.7\,GHz (14\,mm), 43.4\,GHz (7\,mm), 86.8\,GHz (3.5\,mm) and 129.3\,GHz (2.3\,mm) since late 2012 and is ongoing. The program has been described in detail by \citet{imo13,imo15} and \citet{carlos15}. The current experiment, the 32nd in the program was performed over 24 hours from March 1, 2016 until March 2, 2016. The observations were observed in left-hand circular polarisation (LCP) mode at a recording rate of 1\,Gbps with approximately 5\,min long scans. The data were recorded with 2\,bit sampling over four intermediate frequencies (IFs), with a total bandwidth of 512\,MHz. The data were correlated at the Korean Astronomical and Space Institute in Daejeon, Korea.

\section{VLBI Data Calibration\label{sec:calibration}}

In general, there are two main stages of calibrating a VLBI dataset: amplitude calibration and phase calibration. These processes have been described in great detail in other sources such as \citet{bible} and also \citet{marti12} for a description relevant to the complications of high frequency VLBI. Here we present a brief overview of the traditional approach to VLBI data reduction. In addition to the amplitude and phase calibration, additional steps of calibration such as bandpass calibration (performed using the AIPS task \emph{BPASS}) and amplitude corrections computing from visibility autocorrelations (using the AIPS task \emph{ACCOR}) are sometimes also performed. 

\subsection{Phase Calibration}\label{sec:phasecal}
Phase calibration has traditionally been the most difficult and most time consuming process in VLBI data reduction. In single polarisation data, there are two main phase calibration steps. The main tool in calibrating phases is a process called fringe fitting, where the amplitudes and phases are Fourier transformed into the delay and rate domain, and the peak in this plane is used to correct residual errors in the delays and rates and is normally performed using the AIPS task \emph{FRING}. The first of the two steps involves the calibration of the phases within the intermediate frequencies (IFs), where unknown delays and phases are introduced by frequency dependent residual delay and phase errors. While this can often be calibrated for by the injection of pulses into the signal path, close to the receiver in VLBI, many stations may not be equipped with such features. Therefore, a process commonly known as \emph{manual phase calibration} (MPC) is performed. MPC is performed by setting the delays and phases to zero of a reference antenna and performing a global fringe-fit (see \citet{fringe} for details about fringe-fitting) to find delay and rate solutions independently for each IF. These solutions are then extrapolated to the rest of the dataset. MPC is typically performed on a strong source which will hopefully have common visibility with all stations in the experiment so that all other stations can be re-referenced against the reference antenna. If this is not the case, multiple MPC steps can be performed, by changing the reference antenna to a station that has already had the MPC applied and then performing MPC on scans with mutual visibility with the new reference antenna and the still uncorrected stations. This method assumes that the instrumental effects do not significantly change over the course of the experiment, which may not be the case and is a significant limitation of MPC. Once the MPC has been performed, a global fringe fit or multi-band solution is then performed.

\subsection{Amplitude Calibration}
Amplitude calibration is complicated at higher frequencies by the rapidly varying absorption effects in the atmosphere, so that in addition to the noise level of the receiver hardware (system temperature or $T_{\textbf{sys}}$), a time dependent opacity correction needs to be included. In a typical VLBI observation, $T_{\textbf{sys}}$ values are provided to the observer. These values must then be formatted in such a way as to be directly readable into AIPS using the AIPS task \emph{ANTAB}. For high frequency observations (at 7\,mm or shorter) it is also important to have station-based weather information provided. Of most importance for amplitude calibration is the atmospheric opacity as a function of frequency ($\tau_{\nu}$). Usually, a linear fit between $T_{\textbf{sys}}$ and airmass (computed from the weather data) is performed and then $T_{\textbf{sys}}$ recomputed assuming zero air-mass \citep[e.g.][]{marti12,bible}. Once these corrections have been applied, the final $T_{\textbf{sys}}$ values are used to convert the data from arbitrary units into Janskys (Jy). This is typically performed using the AIPS Task \emph{APCAL}. In the case of the KVN, the opacity corrections have already been applied to the $T_{\textbf{sys}}$ measurements and do not need to be corrected later in AIPS \citep{kvn}.

\section{The KVN pipeline}\label{sec:KVNpipe}

The KVN pipeline is loosely based on functions from the EVN pipeline (in particular, \emph{evn$\_$funcs.py}). A flow-chart describing the general operation of the pipeline is shown in Fig. \ref{fig:flowchart}. The KVN pipeline implements several additional features compared with a standard AIPS based pipeline such as the EVN pipeline. The first is the automation of the manual phase calibration with the use of the \emph{IF-Aligner} (Section \ref{sec:ifalign}). The second utilises the simultaneous observation capabilities of the KVN to perform a frequency-phase-transfer (FPT) from lower frequencies to higher frequencies (Section \ref{sec:FPT}). These are briefly described below.

\begin{figure*}[htb]
\centering
\includegraphics[width=\textwidth]{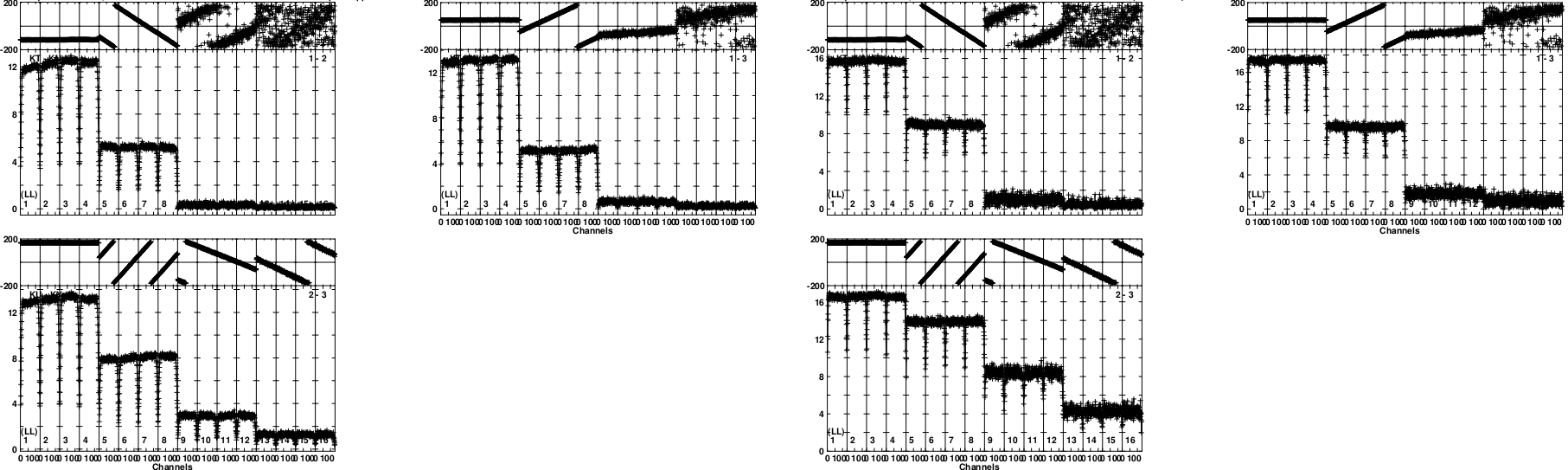}
\caption{Combined \emph{POSSM} plots of all frequencies from the experiment. Left: Uncalibrated data; Right: Amplitude, bandpass and autocorrelation corrected data. Vertical axes is amplitude in Jy (bottom) and phase in degrees (top). \label{fig:combined_uncal}}
\end{figure*}

% \begin{figure*}[htb]
% \centering
% \includegraphics[width=\textwidth]{2mm_if_align_fpt.png}
% \caption{ \emph{POSSM} plots at 2\,mm from the experiment. Left: After IF-Aligner; Right: After FPT \label{fig:ifalign_fpt}}
% \end{figure*}

\begin{figure*}[htb]
\centering
\includegraphics[width=\textwidth]{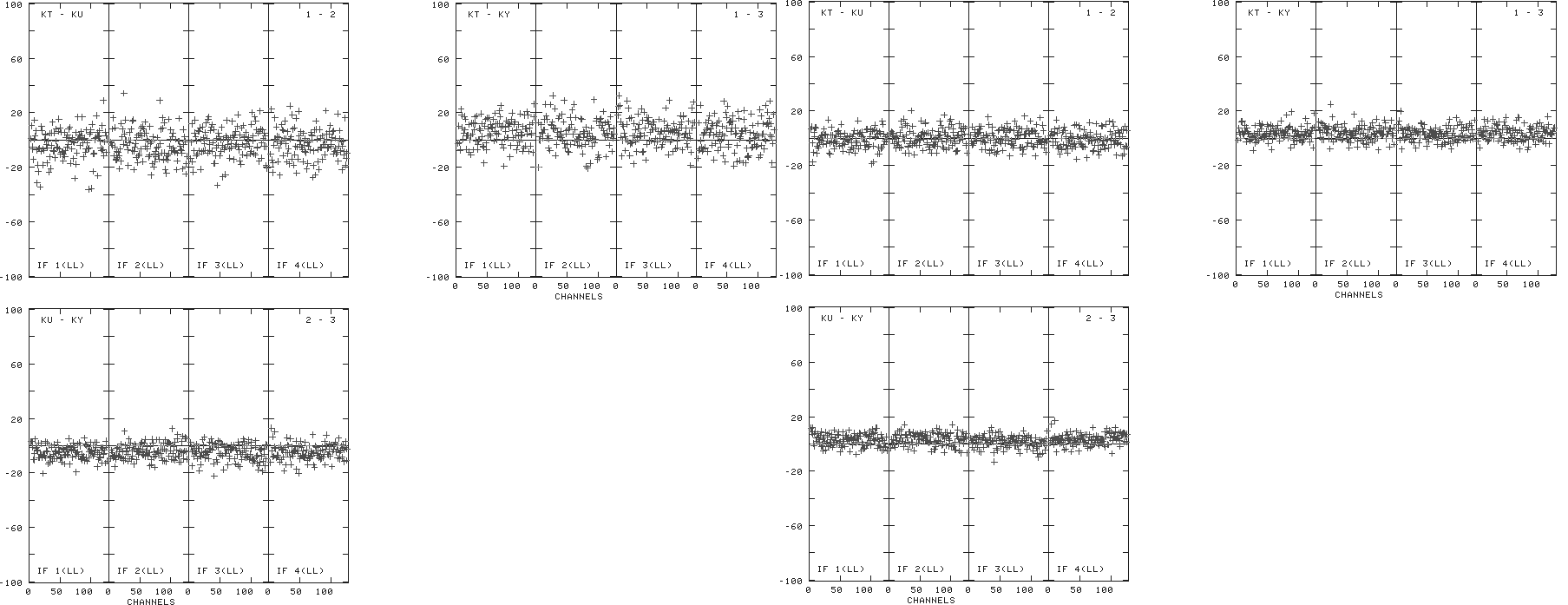}
\caption{Comparison of phases (vertical axes) at 2\,mm between traditional (left) and FPT method (right). Note the reduced phase scatter in the KVN Pipeline version. \label{fig:comparepossm}}
\end{figure*}

\subsection{IF-Aligner}\label{sec:ifalign}

The IF-Aligner is a ParselTongue script described in Section 4.1 of \citet{marti12} that automates and improves upon the standard MPC described in Section \ref{sec:phasecal}. The script works by first performing a global fringe fit, finding independent delays and rates for all data in the observations and then high (greater than 10) signal-to-noise ratio scans are selected. One IF is selected (normally one of the central IFs) and the other IFs are then re-referenced against it. The rate and delay differences in the non-reference IFs are then interpolated over the entire experiment in order to correct weaker scans and then all IFs are re-referenced against the reference antenna. This method has several advantages over using \emph{FRING} to solve for the delays. The first being that a ``good'' scan does not need to be manually selected. The second being that we do not need to assume that the delays are the same at any given frequency band and thirdly, under FRING, the delays are solved for at a given time-range and then applied to the whole experiment, under the assumption that they do not significantly change over the course of the experiment, while the IF-aligner takes time dependent effects into account by interpolating the solutions between good scans. 

\subsection{Frequency Phase Transfer}\label{sec:FPT}

\begin{table*}[t]
\caption{Global fringe detection rates \label{tab:rates}}
\centering
\begin{tabular}{clcccc}
\toprule
Int. time & Type & 2\,mm & 3\,mm & 7\,mm & 15\,mm  \\
\midrule
0.5\,min & Manual (No FPT)   & 16556/4224 (74.4\%)  & 18872/1908 (89.9\%)  & 20372/408 (98.0\%) & 20756/24 (99.9\%) \\
0.5\,min & Manual (FPT)    & 13204/2448 (81.5\%)  & 18004/1704 (90.5\%)  & 20372/408 (98.0\%) & - \\
0.5\,min & Pipeline (FPT)    & 13204/2448 (81.5\%)  & 18004/1704 (90.5\%)  & 20372/408 (98.0\%) & - \\
2.5\,min & Manual (No FPT)   &   3656/516 (85.9\%)  &   4040/132 (96.7\%)  &   4152/20 (99.5\%) &   4172/0 (100\%) \\
2.5\,min & Manual (FPT)    &   3804/368 (90.3\%)  &   4060/112 (97.2\%)  &   4156/16 (99.6\%) & -  \\
2.5\,min & Pipeline (FPT)    &   3804/368 (90.3\%)  &   4060/112 (97.2\%)  &   4156/16 (99.6\%) & -  \\
Scan length & Manual (No FPT)   &   1852/228 (87.6\%)  &    2020/60 (97.0\%)  &    2076/4 (99.8\%) &   2080/0 (100\%) \\
Scan length & Manual (FPT)    &   1956/124 (93.7\%)  &    2032/48 (97.6\%)  &    2076/4 (99.8\%) & - \\
Scan length & Pipeline (FPT)    &   1956/124 (93.7\%)  &    2032/48 (97.6\%)  &    2076/4 (99.8\%) & - \\
\bottomrule
\end{tabular}
\tabnote{A comparison of fringe detection rates at all frequencies both manually reduced with and without FPT and using the KVN Pipeline with the FPT. These are then compared against detection rates at 30 seconds, 2.5\,mins and integrating over the length of a scan (approximately 5\,mins). Numbers are \emph{successful}/\emph{failed} fringe solutions with the percentage of good solutions in brackets. These results were obtained with an SNR limit of 5. }
\end{table*}

Because four wavelengths are recorded truly simultaneously and because atmospheric errors scale predictably with frequency, it is possible to use lower frequency (e.g. 14\,mm) data with longer coherence times to calibrate the phases of higher frequency data (e.g. 2\,mm data with coherence times on the order of seconds). The phases of the lower frequency are simply multiplied by the ratio in the frequencies. This is because the tropospheric component in the phase terms dominates and scales linearly, while other effects are present, these should be removed during the final fringe-fit. To get the best results, there are two competing factors: the number of fringe solutions found which will be determined by the flux density and system temperature (which is typically higher at 14\,mm steep spectrum sources) and the increased non-tropospheric errors due to higher multiplication factors between the frequency bands. In our tests, we find that transferring phases from the most sensitive frequency (normally 14\,mm) compensates for losses due to multiplied non-tropospheric errors. In the case of bad weather or inverted spectrum sources, this may not be the case. As a result of this technique, it is possible to achieve significantly improved sensitivity at higher frequencies. The method is described in greater detail by \citet{rioja11,rioja14a} and \citet{carlos15}.

%While higher frequencies could be used to calibrate further higher frequencies (e.g. 7\,mm used to calibrate 3\,mm), we find that in the most common steep spectrum sources, the additional sensitivity at 14\,mm provides better results than the reduced errors due to smaller frequency ratios.  

\subsection{General Operation}

\begin{table*}[t]
\caption{Image parameter comparison \label{tab:image_parameters}}
\centering
\begin{tabular}{llcccc}
\toprule
Source    &               Type & Peak Intensity & Total flux density & Map noise & Correlated flux  \\
          &                    &     Jy/beam    &     Jy             &  mJy/beam  &    Jy \\
\midrule
3C\,454.3 &   Manual (No FPT)  & 11.17            & 8.71                & 4.32      & $\approx$6.75    \\
3C\,454.3 & Pipeline (FPT)  & 11.17            & 8.78                & 4.33      & $\approx$6.75    \\
%1222+216  &   Manual (No FPT)  & 0.43            & 8.74                & 0.019      & $\approx$0.45    \\
%1222+216  & Pipeline (FPT)  & 0.43            & 8.74                & 0.016      & $\approx$0.45    \\

\bottomrule
\end{tabular}
\tabnote{3C454.3 was examined at 2\,mm in peak intensity, total flux density from model-fitting, map noise and approximate correlated flux on the shortest baseline. }
\end{table*}

\begin{figure*}[htb]
\centering
\includegraphics[width=\textwidth]{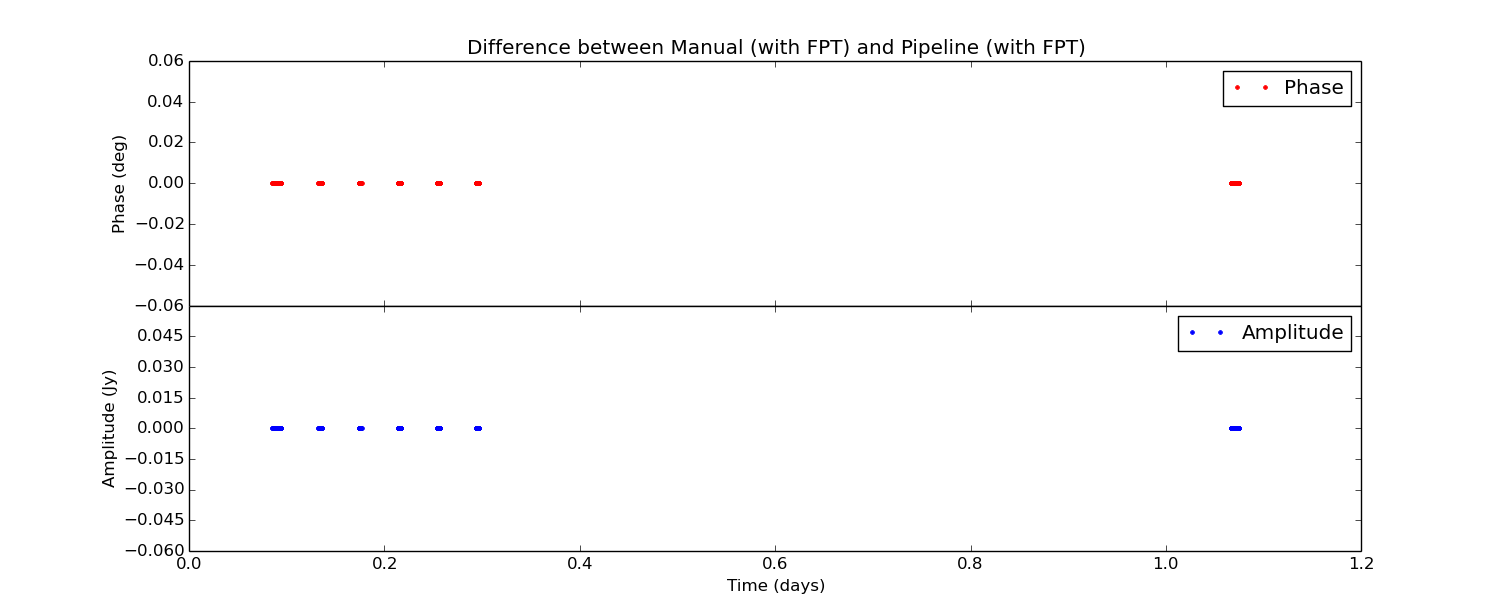}
\caption{The difference between manually reduced (with FPT) and pipelined (also with FPT) phases (top) and amplitudes (bottom) on baseline KTN-KUS for 3C454.3.  \label{fig:vplot_diff}}
\end{figure*}

% \begin{figure*}[htb]
% \centering
% \includegraphics[width=\textwidth]{vplot_mannofpt.png}
% \caption{The phases (top) and amplitudes (bottom) of manually reduced data (with FPT) on baseline KTN-KUS for 3C454.3.  \label{fig:vplot_man}}
% \end{figure*}

% \begin{figure*}[htb]
% \centering
% \includegraphics[width=\textwidth]{vplot_pipefpt.png}
% \caption{The phases (top) and amplitudes (bottom) of pipelined data (with FPT) on baseline KTN-KUS for 3C454.3. \label{fig:vplot_pipe}}
% \end{figure*}

\begin{figure*}[htb]
\centering
\includegraphics[width=\textwidth]{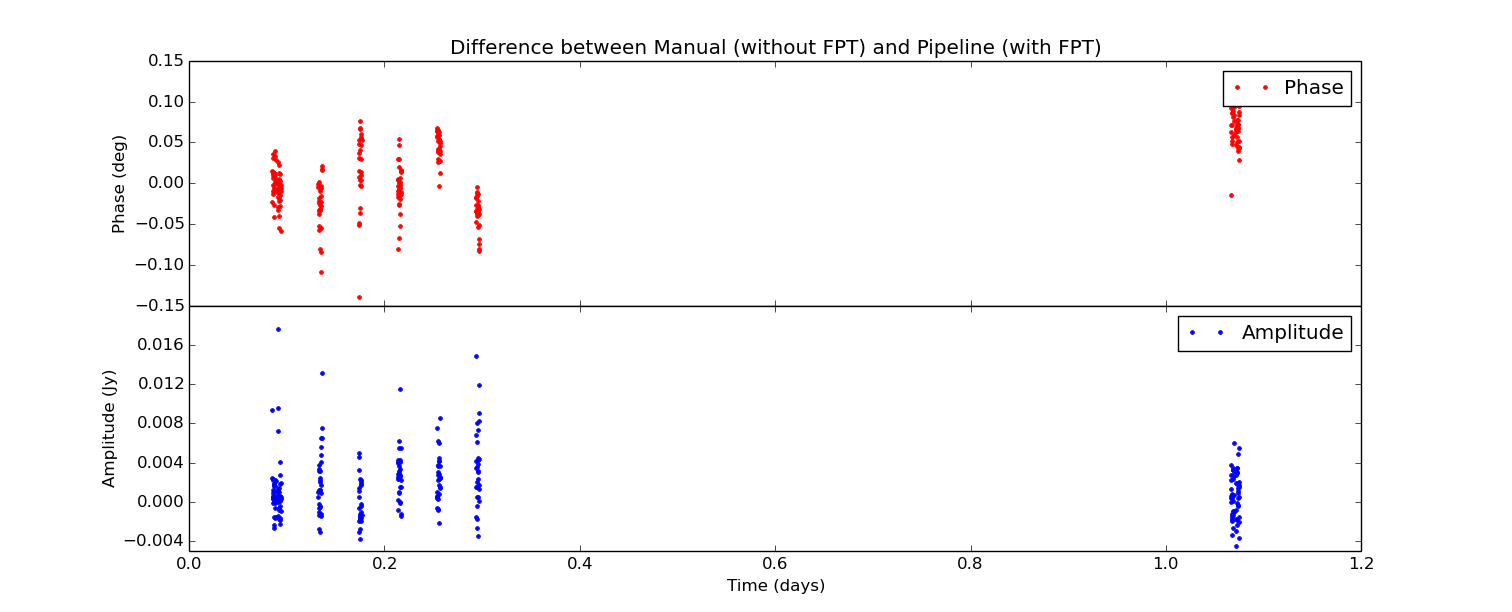}
\caption{The difference between manually reduced (with FPT) and pipelined (also with FPT) phases (top) and amplitudes (bottom) on baseline KTN-KUS for 3C454.3.  \label{fig:vplot_diff2}}
\end{figure*}

\begin{figure*}[htb]
\centering
\includegraphics[width=\textwidth]{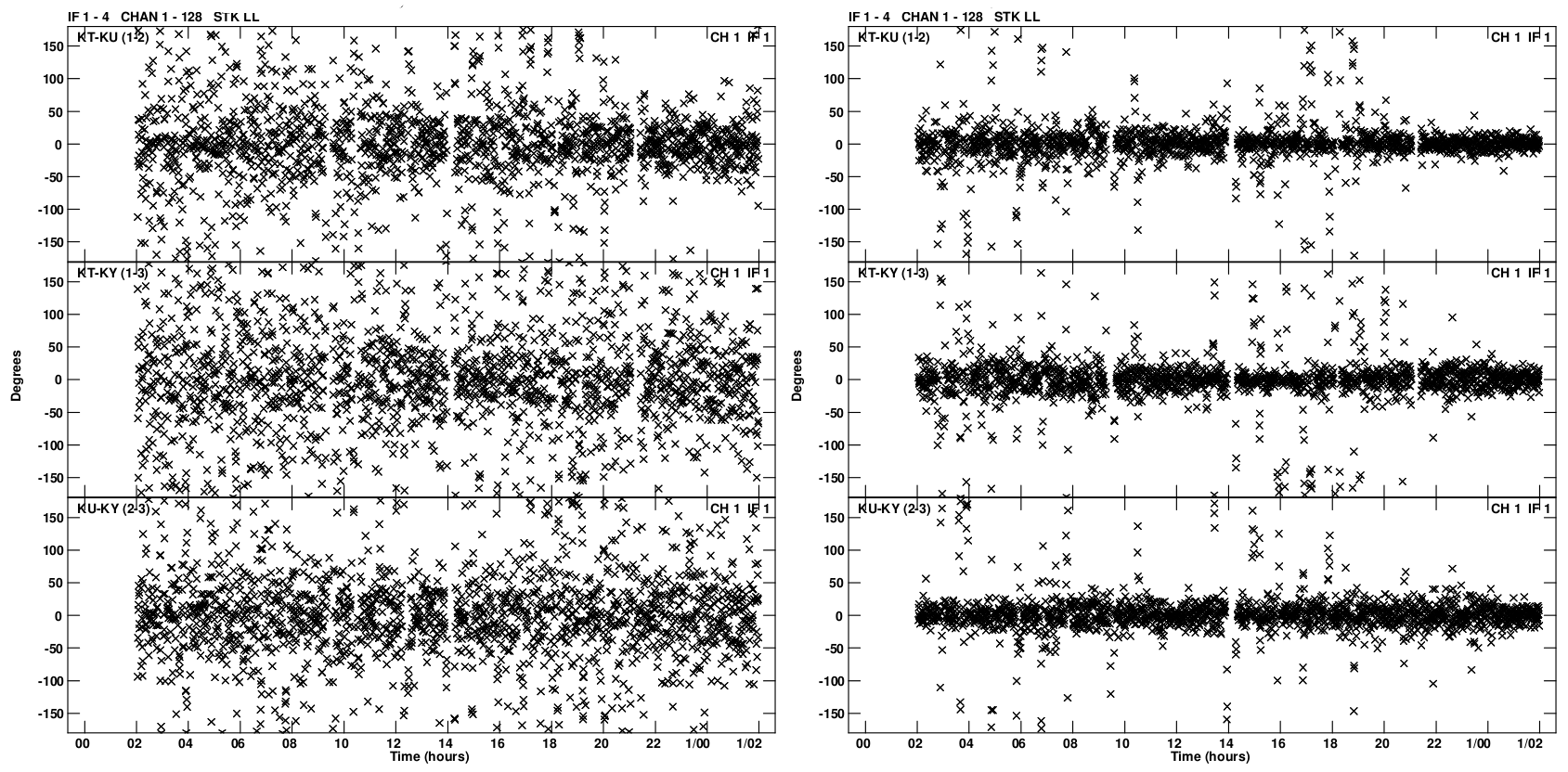}
\caption{Comparison of phase residuals between traditional (left) and FPT method (right).  \label{fig:compare}}
\end{figure*}

Before beginning, the user must provide the raw data from the correlator, \emph{ANTAB}\footnote{We use \emph{ANTAB} files as shorthand for a file containing $T_{\textbf{sys}}$ information for each station.} files for each station in the experiment, and define a location to export the calibrated data. The user must also have AIPS, ParselTongue and Obit installed. In default mode described in Fig. \ref{fig:flowchart}: 

\begin{enumerate}
    \item The script begins by loading the data into AIPS using the AIPS task \emph{FITLD}. A plot of an uncalibrated amplitudes and phases as a function of frequency is presented in Fig. \ref{fig:combined_uncal} (left). The delay errors between the four frequency bands are not related and are not solved for until after separating the dataset by frequency. 
    \item The script then checks the station codes in the \emph{ANTAB} files and changes them to the correct AIPS station codes.
    \item The \emph{ANTAB} files are loaded using the AIPS task \emph{ANTAB} and then applied to the data using the AIPS task \emph{APCAL}. Opacity corrections are not required as they are applied to the $T_{\textbf{sys}}$ values in the \emph{ANTAB} files. A 10\% amplitude correction is also applied at this stage \citep[see:][]{imo15} for 1\,Gbps data.
    \item Amplitude corrections from auto-correlations and amplitude bandpass corrections are then performed using the AIPS tasks \emph{ACCOR} and \emph{BPASS} respectively. A plot showing the amplitudes and phases after these corrections is shown in Fig. \ref{fig:combined_uncal} (right).
    \item The dataset is then separated on the basis of frequency using the AIPS task \emph{UVCOP}.
    \item An initial global fringe-fit and IF-Aligner is run over each separate frequency.
    \item A final global fringe-fit is performed over the 15\,mm dataset with a 30 sec integration time.
    \item The FPT script then transfers the 15\,mm phase solutions to the higher frequency datasets. See Section \ref{sec:FPT}.
    \item The FPT can sometimes introduce some small instrumental errors due to differences in the receivers at the different frequencies. Therefore, a second round of global fringe fitting and IF-aligner can be performed to correct for these errors. This step may make phase referencing less accurate and can be optionally switched off.
    \item A final global fringe fit is performed over the 7\,mm, 3\,mm and 2\,mm datasets, with the solution interval set to the scan length (see Fig. \ref{fig:comparepossm} for a comparison between the KVN Pipeline and manual reduction.).
    \item The final calibrated datasets are then split into separate source files using \emph{SPLIT} and then written to disk using \emph{FITTP}
\end{enumerate}

The KVN Pipeline also automatically generates plots at each step. Additionally, all parts of the KVN Pipeline can be selectively switched on and off and also can be run without performing the FPT. The running time of the pipeline depends on the length of the experiment and the speed of the computer running the pipeline, but typical times for a 24 hour experiment are between 1 and 3 hours.

\section{Discussion}\label{sec:discussion}

In this section, we examine the performance of the KVN Pipeline. In Table \ref{tab:rates}, we compare the fringe detection rates using the traditional reduction procedure described in Section \ref{sec:calibration} against those using the KVN Pipeline using FPT as described in Section \ref{sec:KVNpipe}. Additionally, the dataset was reduced manually using the FPT. In general, both methods produce consistent numbers of good solutions, though in all cases the pipeline using FPT produces a greater numbers of solutions with the manually reduced dataset producing identical number of results as the pipeline. To investigate if the manually reduced (with FPT) dataset is identical to the pipeline version, we subtracted the differences in the visibilities and found the difference to be zero (see Fig. \ref{fig:vplot_diff}).

For this reason, all further comparisons are only between the Manual Non-PFT method of reduction and the KVN Pipeline version. It should however be noted that this experiment had extremely good conditions. We would expect the advantages of FPT to become more apparent if the source cannot be detected at higher frequencies with relatively short integration times. In Fig. \ref{fig:comparepossm} we show \emph{POSSM} plots after the final global fringe-fit using both the manual method (left) and the KVN Pipeline (right). The results are comparable, with differences in amplitude being due to the vector averaging performed in \emph{POSSM}. We can see that the results are very comparable. To further investigate, we subtracted the differences between the amplitudes and phases (Fig. \ref{fig:vplot_diff2}) and find minimal differences, with 0.94$^{\circ}$ difference in phases on average and 0.00016\,Jy difference on average in the amplitudes. Significant improvement can also be seen in the phase residuals (Fig. \ref{fig:compare}), which compares the phase residuals of the manual method (left) and the KVN pipeline (right). 

\begin{figure*}[htb]
\centering
\includegraphics[width=\textwidth]{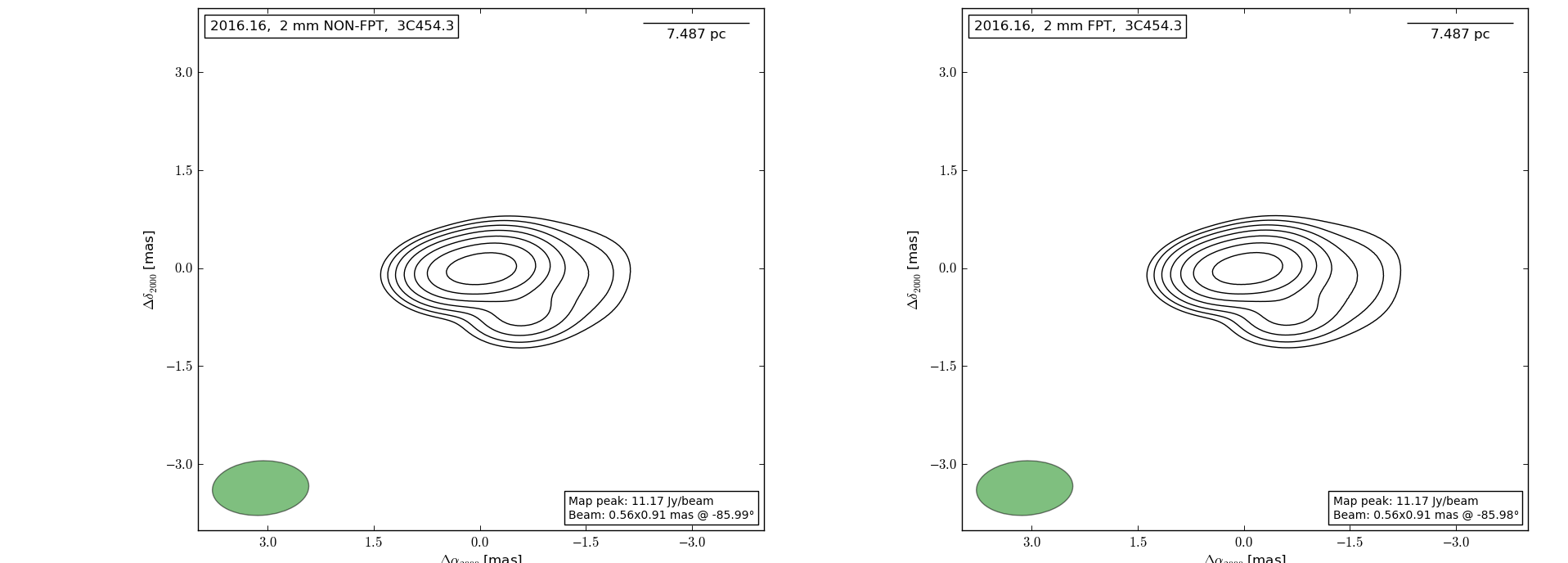}
\caption{ Comparison images between Non-FPT (left) and FPT reduced data (right) in the bright blazar 3C\,454.3. \label{fig:454}}
\end{figure*}

To further check the results of the KVN Pipeline, we imaged and model-fitted a bright source (3C\,454.3) in DIFMAP at 2\,mm and compared the results. The results (displayed in Table \ref{tab:image_parameters}) show that the both methods create very comparable images, with almost identical peak intensities and noise levels. Our testing indicates that for bright sources, there are almost no differences in image quality between the FPT and non-FPT methods. A slightly higher total flux was recovered using the FPT dataset, but the difference is less than 1\%. The CLEAN maps are displayed in Fig. \ref{fig:454}.

\section{Conclusions\label{sec:con}}

The KVN pipeline has demonstrated its ability to automatically reduce a single polarisation KVN datset without direct human intervention. The performance of the KVN Pipeline is consistently superiour to manually reduced data without FPT being applied. While the KVN is a simple network consisting of 3 identical telescopes, the methods described here (although not including the FPT capability) should be applicable to less homogoneous arrays such as the GMVA, KaVA and the EAVN. Currently the pipeline is only capable of reducing single polarisation continuum data. In future versions of the pipeline, we wish to add a polarisation reduction capability, spectral line reduction and possibly automated imaging and phase referencing capabilities.

%%% ACKNOWLEDGMENTS (IF ANY) %%%%%%%%%%%%%%%%%%%%%%%%%%%%%%%%%%%%%%%%

\acknowledgments

The KVN is a facility operated by the Korean Astronomy and Space Science Institute. The KVN operations are supported by KREONET (Korean Research Environment Open NETwork) which is managed and operated by KISTI (Korean Institute of Science and Technology Information).

%%% APPENDICES (IF ANY) %%%%%%%%%%%%%%%%%%%%%%%%%%%%%%%%%%%%%%%%%%%%%

\begin{thebibliography}{}



\bibitem[Algaba et al.(2015)]{carlos15} Algaba, J.-C., Zhao, G.-Y., Lee, S.-S., et al. \,2015, Journal of Korean Astronomical Society, 48, 237 
\bibitem[Clark(1980)]{clean} Clark, B.~G.\ 1980, \aap, 89, 377 
\bibitem[Greisen(2003)]{aips} Greisen, E.~W.\ 2003, Information Handling in Astronomy - Historical Vistas, 285, 109 

\bibitem[Kettenis et al.(2006)]{parsel} Kettenis, M., van Langevelde, H.~J., Reynolds, C., \& Cotton, B.\ 2006, Astronomical Data Analysis Software and Systems XV, 351, 497 
\bibitem[Lee et al.(2013)]{imo13} Lee, S.-S., Han, M., Kang, S., et al.\ 2013, European Physical Journal Web of Conferences, 61, 07007 
\bibitem[Lee et al.(2014)]{kvn} Lee, S.-S., Petrov, L., Byun, D.-Y., et al.\ 2014, \aj, 147, 77 
\bibitem[Lee et al.(2015)]{imo15} Lee, S.-S., Byun, D.-Y., Oh, C.~S., et al.\ 2015, Journal of Korean Astronomical Society, 48, 229 
\bibitem[Mart{\'{\i}}-Vidal et al.(2012)]{marti12} Mart{\'{\i}}-Vidal, I., Krichbaum, T.~P., Marscher, A., et al.\ 2012, \aap, 542, A107 
\bibitem[Pearson \& Readhead(1984)]{fringe} Pearson, T.~J., \& Readhead, A.~C.~S.\ 1984, \araa, 22, 97 
\bibitem[Reynolds et al.(2002)]{evnpipe1} Reynolds, C., Paragi, Z., \& Garrett, M.\ 2002, arXiv:astro-ph/0205118 
\bibitem[Rioja \& Dodson(2011)]{rioja11} Rioja, M., \& Dodson, R.\ 2011, \aj, 141, 114 
\bibitem[Rioja et al.(2014)]{rioja14a} Rioja, M.~J., Dodson, R., Jung, T., et al.\ 2014, \aj, 148, 84 
\bibitem[Taylor et al.(1999)]{bible} Taylor, G.~B., Carilli, C.~L., \& Perley, R.~A.\ 1999, Synthesis Imaging in Radio Astronomy II, 180,  
\bibitem[Zensus \& Ros(2015)]{evn} Zensus, J.~A., \& Ros, E.\ 2015, arXiv:1501.05079 

\end{thebibliography}
\end{document}